\documentclass[12pt]{article}
\usepackage{graphics}
\usepackage{hyperref}
\usepackage{graphicx}
\usepackage{amsmath}
\usepackage{amssymb}
\usepackage{amstext}
\usepackage{epstopdf}
\usepackage{amscd}
\usepackage{amsfonts}
\usepackage{appendix}
\usepackage{color}
\usepackage{wasysym}
\usepackage{bm}
\usepackage{euler}
\newcommand{\nd}{\noindent}

\newcommand{\be}{\begin{equation}}
\newcommand{\ee}{\end{equation}}
\newcommand{\ben}{\begin{eqnarray}}
\newcommand{\een}{\end{eqnarray}}

\title{{{\bf An Entropic Force Schr\"odringer  mechanism for Dark Matter  generation}}}

\author{{\small{A. Plastino$^{1,3,4}$, M. C. Rocca$^{1,2,3}$}}, \\
\small{$^1$ Departamento de F\'{\i}sica,
Universidad Nacional de La Plata,}\\
\small{$^2$ Departamento de Matem\'{a}tica,
Universidad Nacional de La Plata,}\\
\small{$^3$ Consejo Nacional de Investigaciones Cient\'{\i}ficas
y Tecnol\'{o}gicas}\\
\small{(IFLP-CCT-CONICET)-C. C. 727, 1900 La Plata -
Argentina}\\\small{$^4$  SThAR - EPFL, Lausanne, Switzerland}}

\date{\today}

\begin{document}

\maketitle

\begin{abstract}

\nd Theorists of entropic (emergent) gravity put forward that what has been regarded as
unobserved dark matter might instead be  the product of quantum effects that
can be looked at as emergent energy (EE).  Here we
describe a novel Schr\"odringer  mechanism (SM). This SM uncovers the existence of  new quantum 
gravitational states that could be associated to the above mentioned EE.
 This is done  on the basis of the
microscopic Verlinde-like entropic force advanced in [Physica A   {\bf 511}  (2018) 139], that deviates from the Newton's form at extremely short distances.

\nd KEYWORDS: Dark matter, emergent entropic force, Schr\"odringer equation.\\

\end{abstract}

\newpage

\tableofcontents

\newpage

\renewcommand{\theequation}{\arabic{section}.\arabic{equation}}

\section{Introduction}

\subsection{Entropic emergent gravity}

In 2011 Verlinde \cite{verlinde} thought of linking gravity to an entropic force. The ensuing conjecture was later proved  \cite{p1}, in a classical scenario.\vskip 3mm

\nd In reference \cite{verlinde}, gravity  emerges as a consequence of information regarding the positions of material bodies,
combining a thermal gravitation treatment to 't Hooft's holographic principle. Accordingly, gravitation ought to be viewed as  an emergent phenomenon. Such exciting Verlinde's idea received a lot of attention. For
instance, consult \cite{times,libro}. An outstanding review of the statistical mechanics of gravity is that of  Padmanabhan's  \cite{india}.\vskip 2mm

\nd Verlinde's notions gave rise to works on cosmology, the dark energy hypothesis, cosmological acceleration, cosmological inflation, and loop quantum gravity. The pertinent literature is immense \cite{libro}.  Remark on Guseo's work \cite{guseo}. He demonstrated  that the local entropy function, related to a
logistic distribution, is a catenary and vice versa, an invariance that  can
be conjoined with   Verlinde’s conjecture regarding gravity's  emergent origin.
  Guseo puts forward a new  interpretation of the local entropy in a system  \cite{guseo}.
 
\subsection{Dark matter and emergent gravity}

\nd Theorists of entropic (emergent) gravity put forward that what has been regarded as
unobserved dark matter might instead be  the product of quantum effects that
 produce an emergent energy (EE)  \cite{darkVerlinde,libro1,arti}. {\it This EE will be here attributed to the gravitational interaction between two bosons}. \vskip 3mm

\nd   An example for bosons are the axions. Even if any type of boson will do for our mechanism, here we will use axions. The axion is  an as yet undetected particle   
 introduced by Peccei and Quinn in 1977 \cite{PQ} and used to solve the strong CP problem in the Standard Model (SM). 
As it is well-known, the CP (charge conjugation and parity) symmetry is violated by the weak interaction in the SM. The same does not happen with the strong interaction, which poses a dilemma. The solution was found by Peccei and Quinn by hypothesizing a new particle, the axion (see article by  Frank Wilczek \cite{quanta}). With it, the problem of the conservation of CP symmetry in the strong interaction could be solved. It has a very small mass, roughly $1.25$ milli electron volts \cite{nature}. It is assumed that   they can interact only via  gravitation. A posteriori, the axion was postulated as a candidate for solving the problem of dark matter \cite{bergstrom}. Summing up, the axion is a hypothetical elementary particle. Should it exist, it might be regarded as  a possible component of cold dark matter. 

\subsection{Our present goals}

\nd   Verlinde describes gravity as an emergent phenomenon that springs from the quantum entanglement of small bits of space-time information \cite{darkVerlinde}. In our present proposal, the quantum effects that underlie dark matter generation are simpler. They arise out of the gravitational interaction between two axions. Gravitation, regarded a  la Verlinde as an emergent force, differs at very short distances from the Newton's form. The modified gravitation-potential, introduced into the Schr\"odinger equation yields 
quantified states. The energy of the associated ground state yields, as we will see, a considerable amount of energy that, 
via the Einstein's $mc^2$ relation, might generate dark matter.  \vskip 2mm

 \nd In other words, Verlinde's entropic force, proportional to a gradient of the potential energy, is obtained from the entropy. Here, in such a vein,  we rely on a previous study  \cite{boson} that deals with the well known  statistical treatment of quantum ideal gases. We computed there  the above  mentioned derivative  and obtained  a boson-boson gravitational force therefrom. It is of course proportional to $1/r^2$ for distances larger than one micron, but for smaller distances new, more involved contributions emerge. Thus, the ensuing  potential $V(r)$ differs from the Newtonian one at very short distances. We thus write down the Schr\"odringer equation for {\it our new, more involved than Newton's} $V(r)$ and solve it. The new contributions to the gravitation potential generate a lot of {\it unknown till now} quantum gravitational states, to whom we attribute, at least partially, dark matter's origin.
\subsection{Paper's organization}

\nd In Section 2 we review the details of reference  \cite{boson} necessary for constructing the gravity potential $V(r)$ to be employed in the present effort. We also show how to approximate $V(r)$ in order to perform a quantum  analytical treatment of it.
We set $V(r)= \sum\limits_{i=1}^3\, V_i(r)$. In our central argumentation, that of  Section 3, we solve the Schr\"odinger equation separately for these three pieces. The treatment of $V_1(r)$ yields our most important new results. Rough numerical esttimates are given in Sect. 4, while some conclusions are drawn in Section 5.

\section{The Bosonic Quantum Gravitational Potential }
\subsection{The gravitational potential function for bosons}
\setcounter{equation}{0}

\nd As stated above, the gravitation potential $V(r)$ between two bosons of masses $m$ and $M$, respectively,  for an $N-$boson gas, was derived in \cite{boson} via a micro-canonical treatment taken from \cite{lemons}. In this paper we will deal with axions and, of course, $m=M$. Returning to \cite{boson},   
the entropy $S$ for $N$ bosons of total energy $K$ was therein obtained. 
From it one deduces an entropic force $F_e$, that a la Verlinde, is associated to emerging  gravity. The associated 
 boson-boson potential $V(r)$ \cite{boson} will be discussed in this Section, specializing it later on for axions.

\nd In deriving $V(r)$ in   \cite{boson} one defines two constants,  
$a$ and $b$, for $N$ bosons and total energy $K$, in the fashion ($k_B$ Boltzmann's constant)

\be \label{one} a=(3N)^{\frac {5} {2}}h^3; \hskip 2cm b=32\pi(\pi e mK)^{\frac {3} {2}}, \ee
 together with the relation that defines the proportionality constant  $\lambda$   between $F_e$ and the entropic gradient \cite{boson} ($G$ is  gravitation's constant)   
\be \label{two} \lambda  =8\pi GmM/3Nk_B. \ee It is then shown in \cite{boson} that  $V(r)$ acquires the form

\[V(r)=GmM \frac {b} {a}\left\{\frac {r^2} {2}
\ln\left(1+\frac {a} {br^3}\right)-
\frac {a^{\frac {2} {3}}} {2b^{\frac {2} {3}}}\left\{\frac {1} {2}\ln
\left[\frac {\left[r+\left(\frac {a} {b}\right)^{\frac {1} {3}}\right]^2} 
{r^2-\left(\frac {a} {b}\right)^{\frac {1} {3}}r+\left(\frac {a} {b}\right)^{\frac {2} {3}}}
\right]\right.\right.+\]
\begin{equation}
\label{eq2.1}
\left.\left.\sqrt{3}\left[\frac {\pi} {2}-
\arctan\left[\frac{2r-\left(\frac {a} {b}\right)^{\frac {1} {3}}}
{\sqrt{3}\left(\frac {a} {b}\right)^{\frac {1} {3}}}\right]\right]
\right\}\right\}.
\end{equation}
In this paper we consider an ideal gas of axions, whose number is very great, of the order of $N=10^{79}$ (see Section 4).   $K$ is assumed here to be the amount of energy equivalent to the total  dark matter 
mass in the observable Universe, estimated as $K=2.96\times 10^{84}$ eV \cite{rocca}.

\subsection{A Taylor approximation to $V(r)$}
Schr\"odinger's equation with such an awful potential is obviously not amenable to analytic treatment. Since here we are interested in deriving a dark matter generating {\it mechanism}, we need that kind of treatment, which motivates us to find a
 suitable approximation to $V(r)$. We try an approach that consists in  writing 

\begin{equation}
\label{eq2.2}
V(r)\approx V_1(r)+V_2(r)+V_3(r),
\end{equation}
with  $V_1$ the first order Taylor approach for $r$ small enough.  We do this for $0<r<r_0$, with $r_0=10^{-10}$m.\normalcolor
\begin{equation}
\label{eq2.3}
V_1(r)=\frac {\pi GmM} {6}\left(\frac {b} {a}\right)^{\frac {1} {3}}H(r_0-r)=V_0H(r_0-r_1), 
\end{equation}
with $r_1= 25.0$ micron, an empirical figure \cite{shortest}, the minimum distance at which Newton's force that has been verified to work.
The pertinent approximation for large $r$ has been  obtained in \cite{boson} ($H$ is Heavyside's function and $r_1$ a suitable fixed $r-$value, that we determine a few lines below)
\begin{equation}
\label{eq2.4}
V_3(r)=-\frac {GmM} {r}H(r-r_1).
\end{equation}
For intermediate $r-$values,  $r_0<r<r_1$ (as stated above, there is experimental evidence to choose $r_1=25$ micrometers \cite{shortest}). 
 We call $W(r)$  the harmonic interpolating-form between the two fixed distance values  $r_1 - r_0$. Thus,
\begin{equation}
\label{eq2.5}
V_2(r)=W(r)
\end{equation}
We pass now to a Schr\"odinger  treatment of our approximate potential function for gravity.

\setcounter{equation}{0}

\section{Solution of the Schr\"odinger Equation for bosons}

We have now to deal with   
\begin{equation}
\label{eq3.1}
U''(r)+\left[-\frac {l(l+1)} {r^2}-\frac {2\mu_r} {\hbar^2}V(r)+\frac {2\mu_r} {\hbar^2}E\right]U(r)=0,
\end{equation}
for 

\begin{equation}
\label{eq2.22}
V(r)\approx V_1(r)+V_2(r)+V_3(r).
\end{equation}
We subdivide the treatment into three parts.

\subsection{$V_1$  treatment}

For the first component,  $V_1(r)$, essential for our present purposes, we have (see (\ref{eq2.3}) for $V_0$)

\begin{equation}
\label{eq3.2}
U_1''(r)+\left[-\frac {l(l+1)} {r^2}+\frac {2\mu_r} {\hbar^2}(E-V_0)\right]U_1(r)=0.
\end{equation}
We choose here $E>V_0$. The choice will be proved appropriate a posteriori, in Sects. 3 and 4.
 Define  $s=\sqrt{\frac {8\mu_r(E-V_0)} {\hbar^2}}r$ and the associated solution reads

\begin{equation}
\label{eq3.3}
U_{l1}(r)=A(-is)^{l+1}e^{-is}\phi\left(l+1,2l+2;-is\right)-
B(is)^{l+1}e^{is}\phi\left(l+1,2l+2;is\right),
\end{equation}
where $\phi$ stands for the  hypergeometric confluent function  \cite{gra}. Thus, the radial solution is

\begin{equation}
\label{eq3.4}
R_{l1}(r)=A(-is)^{l+1}\frac {e^{-is}} {r}\phi\left(l+1,2l+2;-is\right)-
B(is)^{l+1}\frac {e^{is}} {r}\phi\left(l+1,2l+2;is\right).
\end{equation}
Consulting \cite{gra} we see that 
\begin{equation}
\label{eq3.5}
\phi(l+1,2l+2;-is)=2^{2l+1}e^{-i\pi(l+\frac {1} {2})}\Gamma\left(l+\frac {3} {2}\right)
s^{-\left(l+\frac {1} {2}\right)}e^{-i\frac {s} {2}}{\cal J}_{l+\frac {1} {2}}\left(\frac {s} {2}\right),
\end{equation}
and then  \cite{gra} 
\begin{equation}
\label{eq3.6}
R_{l1}(r)=2^{2l+1}\Gamma\left(l+\frac {3} {2}\right)\frac{s^{\frac {1} {2}}} {r}\left(
Be^{\frac {3\pi il}{2}}e^{{3is} {2}}-Ae^{-\frac {3\pi il}{2}}e^{-{3is} {2}}\right)
{\cal J}_{l+\frac {1} {2}}\left(\frac {s} {2}\right),
\end{equation}
where  $A$ and  $B$ are two arbitrary constants, and ${\cal J}_{l+\frac {1} {2}}$
is the well-known Bessel function  \cite{gra}.     Note that  
 $R_l$ must comply with 
$R_l(r_0)=0, R_l^{'}(r_0)=0$. Let  $s_0=\sqrt{\frac {8\mu_r(E-V_0)} {\hbar^2}}r_0$. Then, the two boundary conditions lead to 
 the demand
\begin{equation}
\label{eq3.7}
{\cal J}_{l+\frac {1} {2}}\left(\frac {s_0} {2}\right)=0.
\end{equation}
Accordingly, $s_0/2$ must be a zero of  the Bessel function. This zero will be called $\chi_{l,n}$, i.e., 
\begin{equation}
\label{eq3.8}
s_0=2\chi_{l,n} = \sqrt{\frac {8\mu_r(E-V_0)} {\hbar^2}}r_0.
\end{equation}
From here we see that the eigenvalue $E$ becomes a {\it quantized energy}
\begin{equation}
\label{eq3.9}
E_{l,n}=\frac {\hbar^2} {2\mu_r}\frac {\chi_{l,n}^2} {r_0^2}+V_0,
\end{equation}
where the zeros of the Bessel function provide the quantization scheme. We will see below in Sect. 4  that $V_0$ is very small. 
Thus, given the numerical value of the Bessel zeroes \cite{gra}, we verify now that the choice $E>V_0$ was appropriate. 
With the  definitions (2.1) for $a$ and $b$, and the considerations made below that equation,  one has now  
$a^{\frac {1} {3}}=(3N){\frac {5} {6}}h$, $b^{\frac {1} {3}}=(32\pi)^{\frac {1} {3}}(\pi emK)^{\frac {1} {2}}$.
 We have already fixed $K$ but not yet $N$. For $N$ we have given above just an order of magnitude. It will be 
further assessed in Sect. 4.

\subsection{$V_2$  treatment}

For the interpolating potential we have 
\begin{equation}
\label{eq3.16}
U_2''(r)+\left[-\frac {l(l+1)} {r^2}+\frac {2\mu_r} {\hbar^2}(E-W(r))\right]U_2(r)=0.
\end{equation}
Here we must comply with four boundary conditions  
$R_{l2}(r_0)=0, R_{l2}^{'}(r_0)=0$,   and   $R_{l2}(r_1)=0, R_{l2}^{'}(r_1)=0$. 
Since we have two arbitrary constants and the boundary conditions are four, we can only satisfy two of the four boundary conditions. The first two boundary conditions are satisfied by 
one of the arbitrary constants and by the energy-value.  The third
contour condition should cancel the other arbitrary constant and, therefore, makes the pertinent solution to vanish.
This entails $R_{l2}(r)=0$.

\subsection{$V_3$  treatment}

$V_3$ is the Newton gravitation potential. Here we deal with 
\begin{equation}
\label{eq3.17}
U_3''(r)+\left[-\frac {l(l+1)} {r^2}+\frac {2\mu_r} {\hbar^2}\left(E+\frac {GmM} {r}\right)\right]U_3(r)=0,
\end{equation}
noting that Whitaker's function $W$ solves the differential equation
\begin{equation}
\label{eq3.18}
W''+\left(-\frac {1} {4}+\frac {\lambda} {z}+\frac {\frac {1} {4}-\mu^2} {z^2}\right)W=0.
\end{equation}

\subsubsection{ $E<0$}

Define $\mu=l+\frac {1} {2}$ and  $\lambda=\frac {GmM} {\hbar}\sqrt{\frac {\mu_r} {2|E|}}$, and  
$s=\sqrt{\frac {8\mu_r |E|} {\hbar^2}r}$. The solution to  (\ref{eq3.11}) becomes
\begin{equation}
\label{eq3.19}
U_3(r)=AW_{\lambda,\mu}(s)-BW_{-\lambda,\mu}(-s),
\end{equation}
where $W_{\lambda,\mu}(z)$ is given by
\[W_{\lambda,\mu}(z)=\frac {(-1)^{2\mu}z^{\mu+\frac {1} {2}}e^{-\frac {z} {2}}}
{\Gamma\left(\frac {1} {2}-\mu-\lambda\right)
\Gamma\left(\frac {1} {2}+\mu-\lambda\right)}\left\{
\sum\limits_{k=0}^{\infty}\frac {\Gamma\left(k+\mu-\lambda+\frac {1} {2}\right)}
{k!(2\mu+k)!}\right.\otimes\]
\[\left[\psi(k+1)+\psi(2\mu+k+1)-\psi\left(\mu+k-\lambda+\frac {1} {2}\right)-\ln z\right]+\]
\begin{equation}
\label{eq3.20}
\left.(-z)^{-2\mu}\sum\limits_{k=0}^{2\mu-1}\frac {\Gamma\left(2\mu-k\right)
\Gamma\left(k-\mu-\lambda+\frac {1} {2}\right)}
{k!}(-z)^k\right\}
\end{equation}
Here $2\mu+1$ is a natural number and the last sum above must vanish for $\mu=0$. Accordingly, 
\begin{equation}
\label{eq3.21}
R_{l3}(r)=r^{-1}[AW_{\lambda,\mu}(s)-BW_{-\lambda,\mu}(-s)].
\end{equation}
The operating boundary conditions are now  $R_{l3}(r_1)=R_{l3}^{'}(r_1)=0$. They lead to 
 
\begin{equation}
\label{eq3.22}
W_{\lambda,\mu}^{'}(s_1)+\frac {W_{\lambda,\mu}(s_1)}
{W_{\lambda,\mu}(-s_1)}W_{-\lambda,\mu}^{'}(-s_1)=0.
\end{equation}
Let  $\sigma_{l,n}$ be the zeroes of this equation. Then, $s_1$ is one of them. .
\begin{equation}
\label{eq3.23}
s_1=\sigma_{l,n}.
\end{equation}
Note that, from experiment \cite{shortest} we can set $r_1= 25$ micrometers.
As we have  (reasoning as in (3.8)) $s_1=\sqrt{\frac {8\mu_r |E|} {\hbar^2}r_1}$, then
the energy becomes {\it quantized} and given, according to the $s_1$ values,  by 
\begin{equation}
\label{eq3.24}
E_{l,n}=-\frac {\hbar^2} {8\mu_r}\frac {\sigma_{l,n}^2} {r_1^2}.
\end{equation}
 For example, setting $l=0$ we have that the energy differences between two contiguous excited states is of the order of  
$10^{-19}$ Joules. The quantization is thus somewhat fictitious. There is an effectively  continuous energy, as one should expect.\normalcolor

\subsubsection{ $E>0$}

 \nd In this instance we have $\mu=l+\frac {1} {2}$, $\lambda=-i\frac {GmM} {\hbar}\sqrt{\frac {\mu_r} {2E}}$, and 
$s=\sqrt{\frac {8\mu_r E} {\hbar^2}r}$. The ensuing solution is
\begin{equation}
\label{eq3.25}
U_3(r)=AW_{\lambda,\mu}(-is)-BW_{-\lambda,\mu}(is),
\end{equation}
and then 
\begin{equation}
\label{eq3.26}
R_{l3}(r)=r^{-1}[AW_{\lambda,\mu}(-is)-BW_{-\lambda,\mu}(is)].
\end{equation}
Boundary conditions are again $R_{l3}(r_1)=R_{l3}^{'}(r_1)=0$, that translate into 

\begin{equation}
\label{eq3.27}
W_{\lambda,\mu}^{'}(-is_1)+\frac {W_{\lambda,\mu}(-is_1)}
{W_{\lambda,\mu}(is_1)}W_{-\lambda,\mu}^{'}(is_1)=0.
\end{equation}
Let  $\varsigma_{l,n}$ be the zeroes of this equation.  Then,
\begin{equation}
\label{eq3.28}
s_1=\varsigma_{l,1}.
\end{equation}
The energy becomes quantized once again

\begin{equation}
\label{eq3.29}
E_{l,n}=\frac {\hbar^2} {8\mu_r}\frac {\varsigma_{l,n}^2} {r_1^2}.
\end{equation}
  Now, setting $l=0$ we have that the energy differences between two contiguous excited states is of the order of  
$10^{-19}$ Joules. The quantization is thus somewhat fictitious. There is a continuous energy, as one should expect.

\setcounter{equation}{0}

\section{Rough numerical estimates}
\nd We put together some numerical values , and extract some important numerical values..

\begin{itemize}
\item 1) Calling $m_a$ to the axion mass, we have $m_a=1.25$ milli electron volt \cite{nature}. 
\item 2) We saw above that the energy equivalent of the total dark mass in the observable Universe is $K= 2.86\times 10^{84}$ eV 
\cite{rocca}.   
\item 3) We verify now that, indeed, $V_0$ is very small, as had been anticipated in Sect. 3 above. 
Its magnitude is $\frac {8.7} {N^\frac {5} {2}}\times 10^{-23}$ eV
Thus (setting $N=1$), we have $V_0<8.7\times 10^{-23}$ eV, which is negligible compared to the first sum
that appears in  (\ref{eq3.9}),  where we have for the ground state  $E_{0,0}$
a value $\chi_{0,0}=\pi$ and $r_0=10^{-10}$ m.Thus, we immediately get 
$E_{0,0}=9.75\times 10^4$ eV. 
\item 4) Therefore the number $N$  of axions in the observable Universe becomes approximately $2K/E_{0,0} \sim 
  6\times 10^{79}$, if, as  assumed here, the energy $E_{0,0}$ is the sole origin of dark matter. We have finally fixed the value of $N$!
\end{itemize}
\section{Conclusions}

\nd The logic of this paper has been as follows.
\begin{itemize}
\item  We start by accepting Verlinde's suggestion that gravitation emerges from an entropy $S$.
\item For a gas of free bosons we calculated in \cite{boson} 1) $S$, 2) Verlinde's entropic force $F_e$, and from it 3) the gravitation potential $V(r)$. We also found in \cite{boson} that  $V(r)$  deviates from the Newton's form only at extremely short distances. 
\item We have  approximated above $V(r)$ in a suitable fashion so as to obtain analytical solutions for  
the Schr\"odinger equation of potential  $V(r)$.
\item The novelty of our treatment emerges at very short distances (the $V_1$ component of $V(r)$). The low-lying 
Schr\"odinger quantum states provide energy-eigenvalues, not accounted for previously. They yield, via Einstein's relation energy$ =mc^2$, a significant quantity of dark matter, of the order of five times the extant quantity of ordinary matter. As a matter of fact, one can limit oneself to the energy of the ground state of our Schr\"odinger equation to account for the extant amount of dark matter in the observable Universe.
\end{itemize}

\nd Thus, we have devised a dark matter generating mechanism in which the entropic gravity potential \`a la Verlinde, emerging from a gas of axions (given by 
Eq. (\ref{eq2.1})), after suitably approximating it, yields  analytical solutions to the concomitant Schr\" odinger equation. 
With these solutions we obtain  a large quantity of quantized gravitational energy. 
 Rough numerical estimates  provide an arguably  substantial amount of unobserved energy 
that, in Verlinde's spirit, could be regarded as dark matter. Indeed, we can accommodate things so that our 
mechanism might account for
a large fraction of the extant dark matter.
\vskip 2mm

\nd We have here tried to establish  {\it that a Schr\"odinger dark matter generating mechanism may exist}. How effective is it is left for subsequent research. Let us insist: our objective here was to obtain a new mechanism for dark matter generation, and this seems to have been accomplished. The mechanism is simple enough and could supplement, or perhaps replace, the amount of dark matter supplied by entanglement between bit of space-time \cite{libro1,nuevos}.

\end{document}